\documentclass[twocolumn,prl,aps]{revtex4}
\usepackage{amsmath}   
\usepackage{amssymb}
\usepackage{amsfonts}
\usepackage{graphicx}

\begin{document}
\title
{Direct observation of $\Gamma - X$ energy spectrum transition in narrow AlAs quantum wells}

\author{A.~R.~Khisameeva$^{a,b}$, A.~V.~Shchepetilnikov$^{a}$, V.~M.~Muravev$^{a}$,  S.~I.~Gubarev$^{a}$, D.~D.~Frolov$^{a}$, Yu.~A.~Nefyodov$^{a}$, I.~V.~Kukushkin$^{a,c}$, C.~Reichl$^{d}$, L.~Tiemann$^{d}$, W.~Dietsche$^{d}$, W.~Wegscheider$^{d}$}
\affiliation{$^a$ Institute of Solid State Physics, RAS, Chernogolovka, 142432 Russia\\ 
$^b$ Moscow Institute of Physics and Technology, Dolgoprudny, 141700 Russia \\
$~c$ National Research University Higher School of Economics, Laboratory for Condensed Matter Physics, Moscow, 101000 Russia \\
$^d$ Solid State Physics Laboratory, ETH Zurich, Otto-Stern-Weg 1, 8093 Zurich, Switzerland \\}
\date{\today}

\date{\today}

\begin{abstract}
Spectra of magnetoplasma excitations have been investigated in a two-dimensional electron systems in AlAs quantum wells (QWs) of different widths. 
The magnetoplasma spectrum have been found to change profoundly when the quantum well width became thinner than $5.5$~nm, indicating a drastic change in the conduction electron energy spectrum. The transformation can be interpreted in terms of transition from the in-plane strongly anisotropic $X_x - X_y$ valley occupation to the out-of-plane isotropic  $X_z$ valley in the QW plane. Strong enhancement of the cyclotron effective mass over the band value in narrow AlAs QWs is reported.          

\end{abstract}

\maketitle

Studies of plasma excitations play an increasingly important role in modern research on the electrodynamics of two-dimensional electron systems (2DESs)~\cite{Lusakowski:16}. Plasma excitations are usually studied in GaAs/AlGaAs heterostructures with a single-valley isotropic 2DES confined to a GaAs layer~\cite{Allen:77, Allen:83, Kukushkin:03}. Continuous developments in the molecular beam epitaxy and heterostructure design have led to creation of a new class of high-quality 2DESs confined to AlAs quantum wells (QWs)~\cite{Shayegan:06, Shayegan:17}. The AlAs QWs provide an important research opportunity by combining a strong native 2D electrons mass anisotropy with the ability to tune valley occupations~\cite{Shayegan:06}. Despite these unique properties, the plasmon phenomena in AlAs 2DESs have remained largely unexplored.

Bulk AlAs is an indirect band semiconductor with energy minima at the six equivalent $X$-points of the Brillouin zone. These valleys are commonly referred to as $X_x$ valley for the $[100]$ direction, $X_y$ valley for the $[010]$ direction, and $X_z$ valley for the $[001]$ direction. The Fermi surfaces at these $X$ minima are strongly anisotropic in $k$-space and are described by different longitudinal ($m_{\rm l}=1.1 \, m_0$) and transverse ($m_{\rm tr} = 0.2 \, m_0$) effective masses, in contrast to the much lighter and isotropic mass ($m^{\ast} = 0.067 \, m_0$) of electrons in GaAs~\cite{Adache}. The Lande $g$-factor of electrons in bulk AlAs ($g^{\ast} = 2$)~\cite{Dietsche:05, Shchepetilnikov:15} is also much larger in magnitude and of a different sign than in GaAs ($g^{\ast} = -0.44$)~\cite{Weisbuch:77, Shchepetilnikov:13, Devizorova:14}. In wide AlAs quantum wells with a width of more than $W=5.5$~nm grown on GaAs $(001)$ substrates, the electrons fill only the $X_x$ $[100]$ and $X_y$ $[010]$ valleys in the quantum well plane with a strongly anisotropic Fermi contour. This is due to the biaxial compression of the AlAs layer which arises as a result of the difference in the lattice constants of AlAs and AlGaAs forming the heterointerface. Instead, electrons confined in narrow ($W<5.5$~nm) AlAs quantum wells occupy a single out-of-plane $X_z$ conduction-band valley which is isotropic in the QW plane~\cite{Kesteren:89}. In previous experiments, the $\Gamma - X$ energy spectrum transition in narrow AlAs QWs was detected by a drastic change in the effective masses measured via the indirect temperature-dependent magnetotransport method~\cite{Yamada:94, Stadt:96, Shayegan:04, Shayegan:09}. 

In contrast, the most direct and precise method to characterize the Fermi surface of semiconductors --- microwave magnetospectroscopy --- has never been applied to explore narrow AlAs quantum wells~\cite{Dresselhaus}. Although the first experiments on 2DES in wide AlAs quantum wells of $15$~nm width revealed a number of unique features in the spectrum of magnetoplasma excitations, in particular, discovery of a gap in the spectrum of plasma excitations in perfectly symmetric circular samples and nontrivial modification of the  plasmon  spectrum  upon redistribution  of charge  carriers  between  different AlAs valleys~\cite{CR:93, Muravev:15, Khisameeva:JETP}. In this work, we have performed precise microwave magnetospetroscopy measurements on a set of narrow AlAs quatum wells of different widths. We have identified a $\Gamma - X$ energy spectrum transition for QWs thinner than $5.5$~nm. Most importantly, our high-precision measurements have revealed a strong enhancement of 2D electrons cyclotron effective mass compared to the bare bulk values. 

\begin{figure}[t!]
\includegraphics[width=0.47 \textwidth]{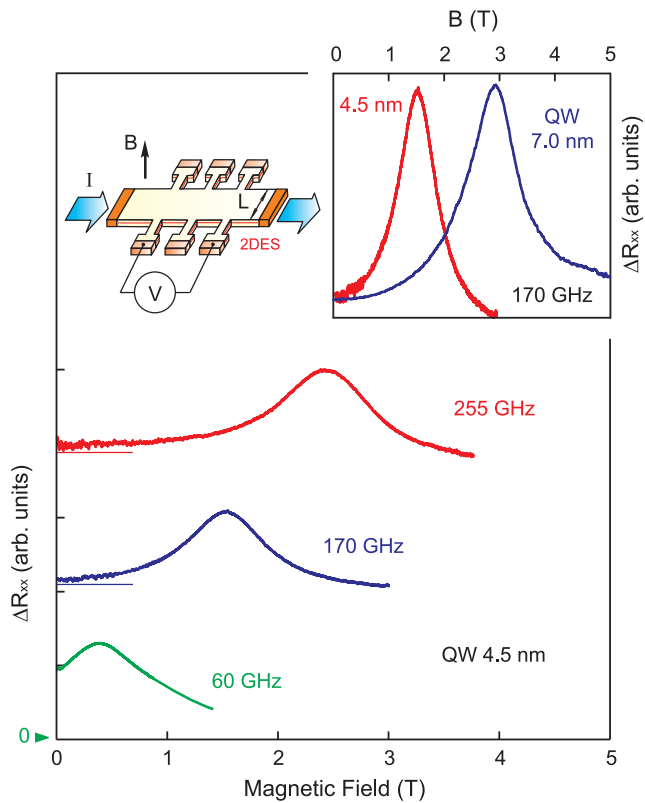}
\caption{ (a) Magnetic-field dependencies of the microwave-induced part of longitudinal resistance $\Delta R_{xx}$ at different microwave frequencies for the sample with QW width $W=4.5$~nm ($n_s = 4.6 \times10^{11}$~cm$^{-2}$). The signal levels without microwaves are denoted by straight lines. The inset shows comparison of $\Delta R_{xx}$ measured at the same microwave frequency $f=170$~GHz for QWs of widths $W=4.5$~nm and $W=7.0$~nm. The magnetoplasmon resonance demonstrates a drastic shift in the magnetic field.}
\label{Fig1}
\end{figure}

The present work is devoted to an experimental study of 2D magnetoplasma excitations in high-quality AlAs/Al$_x$Ga$_{1 - x}$As ($x = 0.46$) quantum wells. The microwave experiments were carried out on AlAs/AlGaAs heterostructures containing $W=4{.}5$~nm, $5{.}5$~nm, or $7{.}0$~nm wide QWs. The heterostructures were grown via molecular beam epitaxy along the $[001]$ direction on an undoped GaAs substrate. The electron density $n_s$ and electron transport mobility $\mu$ at a temperature $T=1.5$~K were in the range of $3{.}5\times 10^{11} - 7{.}0\times 10^{11}\,\text{cm}^{-2}$ and $11 \times 10^{3} - 34 \times 10^{3} \,\text{cm}^{2}/\text{V} \cdot \text{s}$, respectively. The electron density variation was achieved by a short illumination. The sample was patterned into a Hall bar with a width of $L=0.1$~mm and a total length of $2{.}4$~mm (see Fig.~\ref{Fig1} for a schematic drawing). The distance between adjacent potential probes was $1{.}0$~mm. The experimental method was based on strong sensitivity of the longitudinal magnetoresistance $\rho_{xx}(B)$ to the 2DES heating due to resonant microwave absorption when the magnetoplasmon was excited. The measurements were carried out using a double lock-in technique~\cite{Vasiliadou:93}. A sinusoidal probe current of $1~\mu$A at a frequency of $\sim 1$~kHz was driven from a source to a drain contact. The Hall bar was irradiated with amplitude-modulated ($f_{\rm mod} = 31$~Hz) microwaves in the frequency range of $f=60 - 270$~GHz. The first lock-in amplifier detected 2DES resistance $R_{xx}$ between two probe contacts along the Hall bar body. In addition, the microwave-induced part of the longitudinal magnetoresistance $\Delta R_{xx}$ was recorded by a second lock-in connected to the output of the first one. A peak in the $\Delta R_{xx}(B)$ magnetic-field dependence was observed whenever a plasmon was excited in the sample. The sample was placed in the Faraday configuration near the end of a microwave waveguide with a rectangular cross-section of $7.0 \times 3.5$~mm$^2$ (WR~$28$). The experiments were carried out at a sample temperature of $T=(1.5 - 4.2)$~K.



In Fig.~\ref{Fig1} we present the typical dependencies of the microwave-induced part of longitudinal resistance $\Delta R_{xx}$ on the magnetic field $B$ at various irradiation frequencies $f=60$~GHz, $170$~GHz and $255$~GHz. The data were obtained on a sample with density $n_s = 4.6 \times10^{11}$~cm$^{-2}$ and QW width $W=4.5$~nm. Each curve demonstrates a well-defined resonance. The resonant absorption was observed for both signs of the magnetic field. As the microwave frequency $f$ is raised, the resonance shifts to higher magnetic field values and corresponds to the transverse magnetoplasma mode. The inset in  Fig.~\ref{Fig1} shows comparison of $\Delta R_{xx}$ measured at the same microwave frequency $f=170$~GHz for QWs of widths $W=4.5$~nm and $7.0$~nm. Despite the fact that the electron density in the two structures under study is almost identical, there is a drastic change in the resonance magnetic field position, which suggests that the plasma dynamics undergoes a qualitative modification when the QW width is changed from $4.5$~nm to $7.0$~nm. 

\begin{figure}[t!]
\includegraphics[width=0.47 \textwidth]{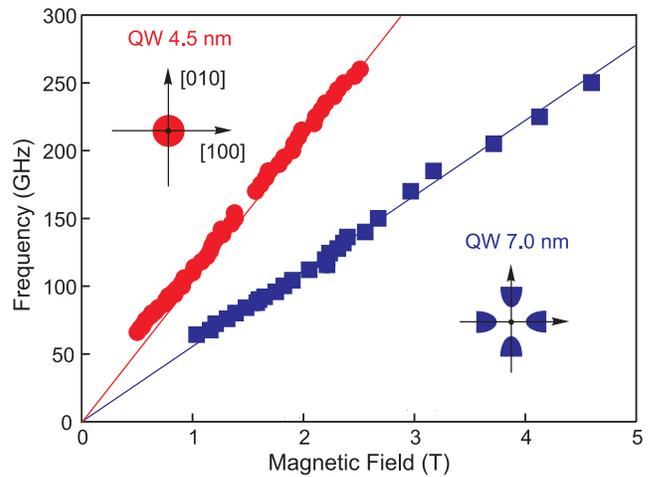}
\caption{Dispersions of two-dimensional magnetoplasma excitations in AlAs quantum wells of width $W=4.5$~nm (red circles) and $W=7.0$~nm (blue squares). The electron density is $n_s = 4.6 \times10^{11}$~cm$^{-2}$. Solid curves mark cyclotron frequency lines as predicted by Eq.~(\ref{eq.1}). 2D Fermi contours for $X_z$ and $X_x - X_y$ valley occupations are schematically presented as insets.}
\label{Fig2}
\end{figure}  

\begin{figure}[t!]
\includegraphics[width=0.47 \textwidth]{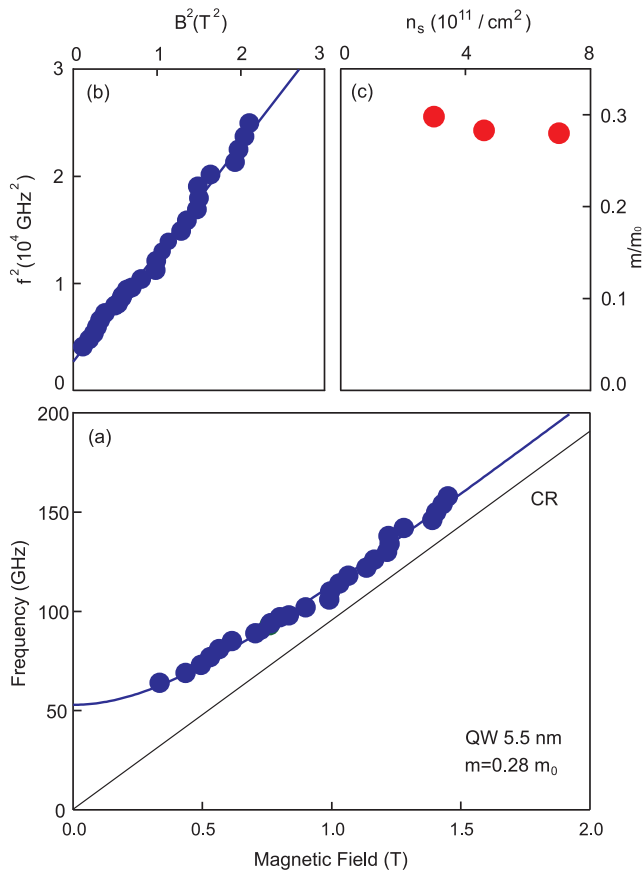}
\caption{ (a) Frequencies of magnetoplasma resonance versus perpendicular magnetic field. The sample electron density is $n_s = 7.0 \times10^{11}$~cm$^{-2}$, the QW width is $W=5.5$~nm. Data points describe the cyclotron magnetoplasma mode. Solid curve is the theoretical prediction from Eq.~(\ref{eq.1}). (b) Magnetodispersion of the plasmon mode in the $f^2 (B^2)$ axes. (c) The electron effective mass $m_c$ as a function of the electron density.}
\label{Fig3}
\end{figure}

In order to gain further insight into the nature of this phenomenon, the experiments of Fig.~\ref{Fig1} were repeated for a wide range of frequencies. Figure~\ref{Fig2} shows magnetic field values at which the magnetoplasmon resonance occurs as a function of the microwave frequency $f$. The measurements were performed for two samples with AlAs QWs widths of $W=4.5$~nm (red dots) and $7.5$~nm (blue squares) which had an identical electron density of $n_s = 4.6 \times10^{11}$~cm$^{-2}$. The resonances in both samples closely follow the cyclotron resonance (CR) line $\omega_c = eB/m_c$ with a minor deviation from linearity in the low-frequency region due to a plasma depolarization effect. The collective magnetoplasma mode frequency is described by the equation~\cite{Chaplik:72}         
\begin{equation}     
\omega^2 = \omega_p^2 + \omega_c^2,
\label{eq.1}
\end{equation} 
where $\omega_c=eB/m_c$ is the cyclotron frequency and  $\omega_p$ is the dimensional plasmon frequency. The cyclotron mass $m_c$ strongly depends on the 2DES valley occupation. If $X_x$ or $X_y$ in-plane valley is occupied, the Fermi contour of the 2D conduction electrons represents an ellipsoid with two effective masses along the main in-plane crystallographic directions $m_{\rm l} = 1.1 \, m_0$ and  $m_{\rm tr} = 0.2 \, m_0$. Hence, the cyclotron mass of anisotropic 2D electrons is determined as a geometric mean of effective masses, $m_c=\sqrt{m_{\rm l} m_{\rm tr}}=0.47\, m_0$.    In contrast, if the 2D electrons occupy the out-of-plane $X_z$ valley, their Fermi contour is reduced to a circle in the QW plane. Thus, the cyclotron mass $m_c$ simply coincides with the value $m = m_{\rm tr} = 0.2 \, m_0$. Fitting Eq.~(\ref{eq.1}) to the experimental data gives the following values of the cyclotron mass $m_c (W=4.5~\rm{nm}) = (0.27 \pm 0.01) \, m_0$, and  $m_c (W=7.0~\rm{nm}) = (0.51 \pm 0.01) \, m_0$. The cyclotron mass obtained for the AlAs QW of width $W=7.0~\rm{nm}$ is almost equal to the value $0.47 \, m_0$ characteristic of the in-plane strongly anisotropic $X_x - X_y$ valley occupation (see the inset in Fig.~\ref{Fig1}). In its turn, the cyclotron mass measured in AlAs QW of width $W=4.5~\rm{nm}$ is close to the conduction band value $m_{\rm tr} = 0.2 \, m_0$, indicating a single out-of-plane isotropic $X_z$ valley occupation~\cite{Kesteren:89, Yamada:94, Stadt:96, Shayegan:04}. The $\Gamma - X$ energy spectrum transition observed resides on a unique physical property of narrow AlAs quantum wells. In thin QWs the energy of spatial quantization dominates over the strain energy, and the electrons tend to fill the valley with the lowest level of spatial quantization, namely, the out-of-plane $X_z$ valley with the largest effective mass in the growth direction $[001]$. For larger $W$, the lattice mismatch between AlAs and GaAs leads to such  strong biaxial compression of the AlAs layer that the electrons start to occupy only the in-plane $X_x$ and $X_y$ valleys.

Note that the obtained value of the $X_z$ valley isotropic mass $m (W=4.5~\rm{nm})=0.27 \, m_0$ is well above the tabulated AlAs band value of $m_{\rm tr} = 0.2 \, m_0$~\cite{Adache}. The same bizarre result was obtained in a number of independent transport experiments~\cite{Yamada:94, Stadt:96, Shayegan:04, Shayegan:09}. To unveil the origin of this discrepancy, we continued our research on the structure with a QW width of $W=5.5~\rm{nm}$ at different electron densities. Fig.~\ref{Fig3}(a) shows the magnetodispersion of the cyclotron magnetoplasma mode excited across the Hall bar with a width of $L=0.1$~mm. The data were obtained at an electron density of $7{.}0\times 10^{11}\,\text{cm}^{-2}$. When fitting the magnetic-field dependence described by Eq.~(\ref{eq.1}) to the experimental data (solid curve in Fig.~\ref{Fig3}(a)), the mode extrapolates to the following zero-field plasma frequency $f_p=\omega_p/2 \pi = 53$~GHz, and cyclotron mass $m_c  = (0.28 \pm 0.01) \, m_0$ (Fig.~\ref{Fig3}(b)). The cyclotron mass value indicates that the isotropic $X_z$ valley is still occupied for the $5.5$~nm AlAs quantum wells. Here, we have an opportunity to compare the effective electron mass obtained from a zero-field plasma frequency and a cyclotron mass. Indeed, the plasmon frequency at $B=0$~T is described by~\cite{Stern:67}
\begin{equation}     
\omega_{p} = \sqrt{\frac{n_s e^2}{2 m_p \varepsilon_0 \varepsilon^{\ast}} q},
\label{eq.2}
\end{equation}
where $\varepsilon^{\ast}=(\varepsilon_{\rm GaAs}+1)/2$ is the effective dielectric permittivity of the surrounding medium, and $q$ is a plasmon wave vector. For a narrow 2DES strip of width $L$, the plasmon wave vector can be approximately described by $q= \pi N /L$ ($N=1,2, \ldots$). More accurate calculations of the zero field plasma frequency give $0.85 \, \omega_p$~\cite{Mikhailov:05}. Taking the plasmon frequency as $f_p = (53 \pm 3)$~GHz Eq.~(\ref{eq.2}) yields the following value of the plasmon mass $m_p=(0.33 \pm 0.04) \, m_0$. Both the cyclotron $m_c$ and the plasmon $m_p$ masses significantly exceed the AlAs  band value of $m_{\rm tr} = 0.2 \, m_0$.     

To further explore the nature of the phenomenon discovered, we performed the experiment of Fig.~\ref{Fig3}(a) for a set of 2D electron densities $n_s=3{.}0\times 10^{11}$, $4{.}6\times 10^{11}$, and $7{.}0 \times 10^{11}\,\text{cm}^{-2}$ ($W=5.5~\rm{nm}$). For more $f^2$ versus $B^2$ data at different electron density levels we refer to Fig.~S1 in the Supplemental Material~\textrm{I} ~\cite{Supplemental}. The electron density was varied by briefly illuminating the sample with a light emitting diode. The experimental dependence of the electron effective mass on the electron density is shown in Fig.~\ref{Fig3}(c). The main result is that the cyclotron mass is almost independent of the 2D density $n_s$. This finding excludes non-parabolicity and retardation effects as a possible reason for the mass increase~\cite{Hopkins:87, Kukushkin:03}. For the detailed analysis of the non-parabolicity effect we refer to Supplemental Material~\textrm{II}~\cite{Supplemental}. The phenomenon discovered is of great physical significance, and will be a focus of our future research. We can speculate that Coulomb correlations may play an important role for mass renormalization in narrow QWs~\cite{Shayegan:09, Vankov:15}. We cannot forget, however, the Kohn theorem, which states that the cyclotron effective mass should be independent of the short-range interaction~\cite{Kohn, Kukushkin}. Also, the AlGaAs barrier band structure may have an effect on the 2D electrons effective mass in the quantum well. More experiments are certainly imperative in order to be able to choose the right hypotheses.       

In summary, we have performed highly accurate magnetospectroscopy experiments on AlAs/Al$_x$Ga$_{1 - x}$As ($x = 0.46$) quantum wells of different widths from $4.5$~nm to $7.0$~nm. The cyclotron mass analysis has revealed a $\Gamma - X$ energy spectrum transition in AlAs quantum wells thinner than $5.5$~nm. Remarkably, we have observed a strong enhancement of the cyclotron effective mass over the band value in narrow AlAs quantum wells. Additional experiments have ruled out non-parabolicity and retardation effects as a possible reason for the mass increase.

The authors gratefully acknowledge financial support from the Russian Science Foundation (Grant No.~14-12-00693).

\end{document}


\title{Supplementary Material for \\ Direct observation of $\Gamma - X$ energy spectrum transition in narrow AlAs quantum wells}

\author{A.~R.~Khisameeva$^{a,b}$, A.~V.~Shchepetilnikov$^{a}$, V.~M.~Muravev$^{a}$,  S.~I.~Gubarev$^{a}$, D.~D.~Frolov$^{a}$, Yu.~A.~Nefyodov$^{a}$, I.~V.~Kukushkin$^{a,c}$, L.~Tiemann$^{d}$, W.~Dietsche$^{d}$, C.~Reichl$^{d}$, W.~Wegscheider$^{d}$}
\affiliation{$^a$ Institute of Solid State Physics, RAS, Chernogolovka, 142432 Russia\\ 
$^b$ Moscow Institute of Physics and Technology, Dolgoprudny, 141700 Russia \\
$~c$ National Research University Higher School of Economics, Laboratory for Condensed Matter Physics, Moscow, 101000 Russia \\
$^d$ Solid State Physics Laboratory, ETH Zurich, Otto-Stern-Weg 1, 8093 Zurich, Switzerland \\}

\date{\today}\maketitle

\section{\textrm{I}. DEPENDENCE OF ELECTRON EFFECTIVE MASS ON ELECTRON DENSITY}

\begin{figure}[!h]
\includegraphics[width=0.47 \textwidth]{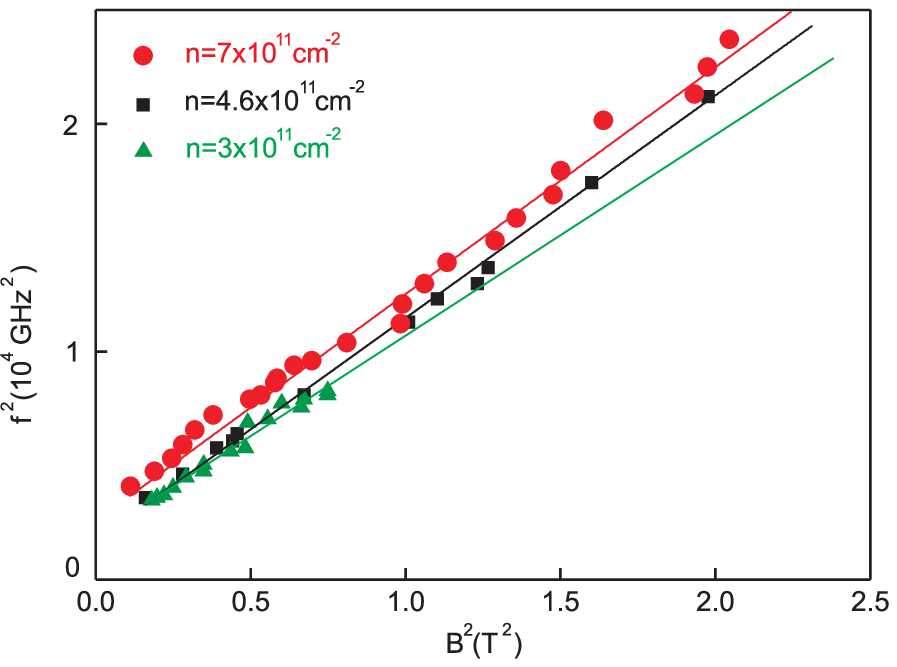}
\caption{Square of the magnetoplasma frequency versus square of magnetic field. The sample electron densities are $n_s=3{.}0\times 10^{11}$, $4{.}6\times 10^{11}$, and $7{.}0 \times 10^{11}\,\text{cm}^{-2}$}. The inset shows electron effective mass $m_c$ as a function of the electron density.
\label{f1}
\end{figure}

In order to explore the nature of the effective mass increase, we performed the experiment on the structure with a QW width of $W=5.5~\rm{nm}$ at different electron densities. The effective mass is best identified by plotting the square of resonant magnetic field versus square of microwave frequencies as shown in Fig.~\ref{f1}. The data were obtained at electron densities of $n_s=3{.}0\times 10^{11}$, $4{.}6\times 10^{11}$, and $7{.}0 \times 10^{11}\,\text{cm}^{-2}$. The electron density was varied by briefly illuminating the sample with a light emitting diode. The values of effective masses for different 2D densities are collected in the inset of Fig.~\ref{f1}. According to obtained results, the cyclotron mass $m_c  = (0.28 \pm 0.01) \, m_0$ was almost independent of the 2D density $n_s$.

\section{\textrm{II}. ANALYSIS OF THE MASS INCREASE}

For polar semiconductors non-parabolicity can lead to modification of the band structure and an increase of the effective mass due to rise of electron density. In 2DES the experimental dependence of cyclotron effective mass on electron concentration were observed earlier in AlGaAs/GaAs heterostructures~\cite{Hopkins:87}. The energy $E$ in quasi-2DES is determined by Fermi energy and the spatial quantization $E_{dim}$.  


The contribution of spatial quantization energy is estimated as  
\begin{equation}
\label{Ed}
E_{dim}=\frac{\pi^{2}\hbar^{2}}{2m^{\ast}W^2} = 11.2 meV
\end{equation}
where $m^{\ast} = 1.1 \, m_0$ and $W=5.5~\rm{nm}$.

The Fermi energy for dis described by the equation
\begin{equation}
\label{Ef1}
E_{F}(n_s = 3{.}0\times 10^{11})=\frac{\pi^{2}\hbar^{2} n_{s}}{m^{\ast}} = 4.1 meV
\end{equation}

\begin{equation}
\label{Ef2}
E_{F} (n_s = 7{.}0\times 10^{11}) = 9.7 meV
\end{equation}
where $m^{\ast} = 0.2 \, m_0$. 

Simply add this two parts together we obtain the changing in energy $\Delta E\approx 15-20 meV$. According to~\cite{book}, the such changing $\Delta E$ should lead to effective mass increase $\Delta m \approx 0.002 \, m_0$. This finding excludes nonparabolicity as a possible reason for the mass increase from $m^{\ast} = 0.2 \, m_0$ to $m^{\ast} = 0.28 \, m_0$ determined from the experiments.